\documentclass[preprint,3p]{elsarticle}
\pdfoutput=1
\usepackage{amsmath,amsfonts,float}
\usepackage{graphicx}
\usepackage{epstopdf}

\begin{document}
	
	\title{Shear to longitudinal mode conversion via second harmonic generation in a two-dimensional microscale granular crystal}
	
	\author[uw]{S. P. Wallen}
	\author[uw]{N. Boechler}

	\address[uw]{ 
		Department of Mechanical Engineering, University of Washington, Seattle, WA, 98195 USA 
	}
	

	\begin{abstract}
		Shear to longitudinal mode conversion via second harmonic generation is studied theoretically and computationally for plane waves in a two-dimensional, adhesive, hexagonally close-packed microscale granular medium. The model includes translational and rotational degrees of freedom, as well as normal and shear contact interactions. We consider fundamental frequency plane waves in all three linear modes, which have infinite spatial extent and travel in one of the high-symmetry crystal directions. The generated second harmonic waves are longitudinal for all cases. For the lower transverse-rotational mode, an analytical expression for the second harmonic amplitude, which is derived using a successive approximations approach, reveals the presence of particular resonant and antiresonant wave numbers, the latter of which is prohibited if rotations are not included in the model. By simulating a lattice with adhesive contact force laws, we study the effectiveness of the theoretical analysis for non-resonant, resonant, and antiresonant cases. This work is suitable for the analysis of microscale and statically compressed macroscale granular media, and should inspire future studies on nonlinear two- and three-dimensional granular systems in which interparticle shear coupling and particle rotations play a significant role.
	\end{abstract}
	
	\maketitle
	
	\section{Introduction}

	Granular media are known to exhibit complex dynamical behavior that stems from their discrete, often heterogeneous, structure and highly nonlinear particulate interactions \cite{DuranBook, GranularPhysicsBook, NesterenkoBook}. Ordered and reduced-dimensional granular systems, often referred to as ``granular crystals,'' have been a setting of interest for several decades, as they have yielded new, broader understanding of granular media dynamics \cite{NesterenkoBook}. In addition, utilization of the nonlinear particulate interactions in conjunction with classical linear effects such as dispersion induced by structural periodicity has resulted in granular crystals being used in new strategies for passive wave tailoring \cite{GranularCrystalReviewChapter,PhysTodayReview}. 
	
	Models used to describe granular crystals are typically composed of one- to three-dimensional (1D and 3D, respectively) systems of rigid bodies interacting via Hertzian normal contact ``springs'' \cite{Hertz,Johnson}. For example, this approach has been used in previous studies to explore highly nonlinear mechanical wave propagation in uncompressed two-dimensional (2D) granular crystals \cite{NesterenkoBook,Shukla1993,Leonard1,Geubelle1,Geubelle2,Leonard2}. While this type of model works well for uncompressed granular media, interparticle shear interactions and particle rotations become increasingly important in statically compressed granular systems. Several recent theoretical works, based on earlier discrete lattice models of elastic media \cite{Suiker2001}, have introduced linear models of two- \cite{Linear2D,ZGVMembrane} and three-dimensional \cite{Linear3DTheory} granular crystals that account for interparticle shear interactions and particle rotations. These studies demonstrated how the additional degrees of freedom and modes of particle coupling can drastically influence the granular crystal dynamics, and yield unique effects such as the rotational waves experimentally observed in statically compressed macroscale granular systems \cite{Linear3DExp}. Shear interactions and particle rotations also play a similarly important role in the emerging field of microscale granular crystals, as has been demonstrated in recent theoretical \cite{WallenPRB} and experimental \cite{HiraiwaPRL} studies of quasi-1D microgranular systems in linear dynamical regimes.  
	
	In this work, we explore the nonlinear phenomena of second harmonic (SH) generation for plane waves traveling in a model of a 2D hexagonally-close packed lattice of microspheres, which includes interparticle adhesive effects, particle rotations, and elastic shear interactions. Second harmonic generation is a well-known phenomenon that has been studied extensively in nonlinear optics \cite{NLOptics} and acoustics \cite{NLAcoustics}. Past works have also examined SH generation in 1D discrete granular chains \cite{Cabaret2012, Sanchez-Morcillo2013} and fluid-saturated granular media \cite{BiotTheory}. Using a successive approximations approach in the manner of Refs. \cite{Cabaret2012, Sanchez-Morcillo2013}, we theoretically analyze the SH generation for cases where the fundamental frequency (FF) wave is purely longitudinal (L) or transverse-rotational (TR) in character, and find that the generated SH waves are longitudinal in both scenarios. Such shear-to-longitudinal mode conversion via SH generation has been previously studied in nonlinear elastic solids \cite{NLSolids}, observed experimentally in 3D granular packings \cite{ModeConversion}, and modeled in quasi-1D nonlinear phononic crystals \cite{GonellaPRL}. We show that the SH generated by a FF wave in the lower TR mode may resonate or vanish for particular wavelengths, but the latter phenomenon is not predicted if rotations are excluded from the model. Finally, we compare our theoretical predictions with dynamic discrete-element simulations of a microscale granular crystal. We find that the theoretical predictions are quantitatively accurate for non-resonant wavelengths, and provide qualitative understanding of the behavior at resonance. This paper extends the already rich body of work on nonlinear waves in 2- and 3D ordered granular media by exploring the interplay of multiple degrees of freedom and nonlinear particulate interactions. 
	
	\section{Theory}
	
	\subsection{Model}
	\begin{figure}[h]
		\centering
		\includegraphics[width=0.75\linewidth]{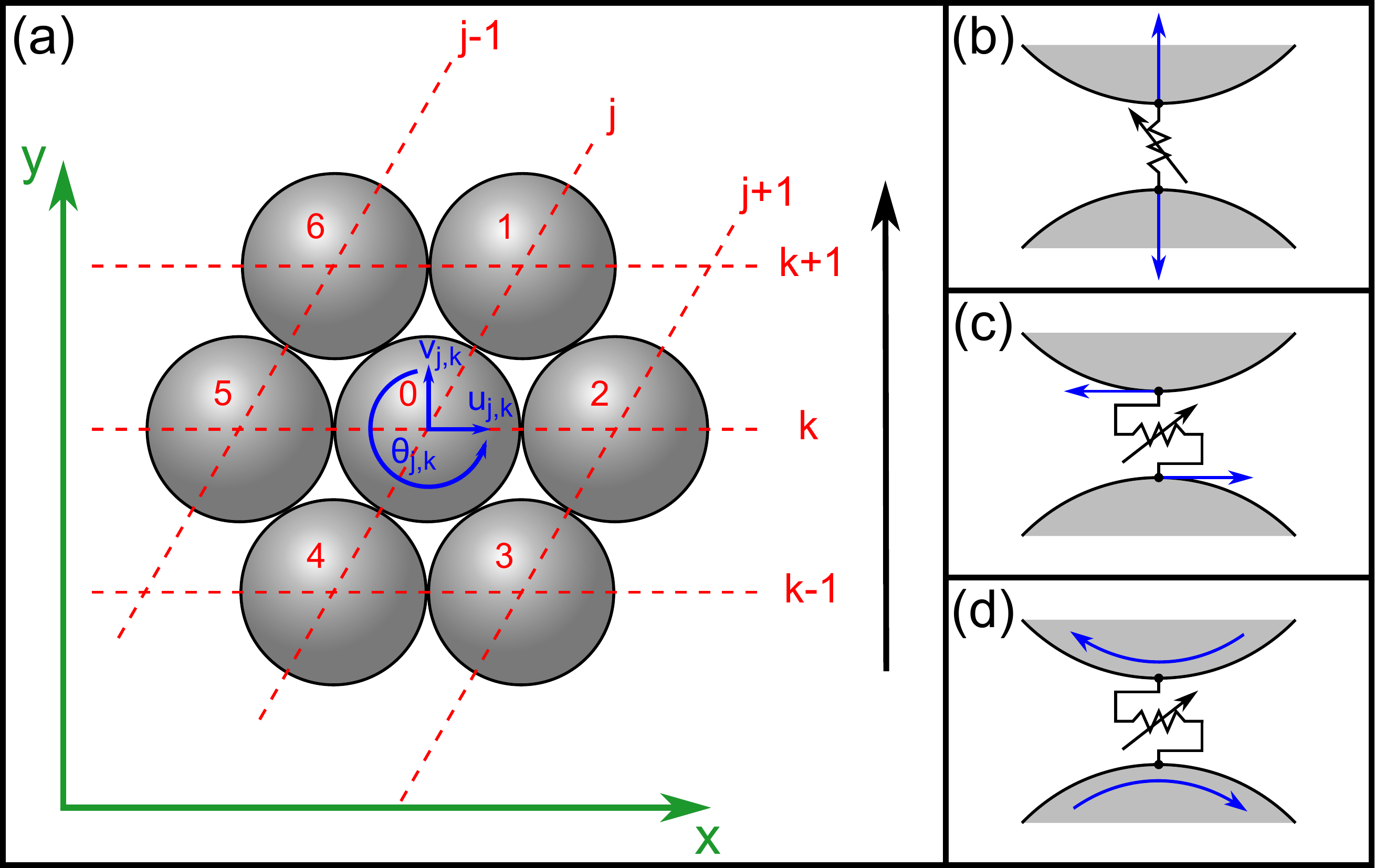}
		\caption{(a) Schematic of the model of a 2D, hexagonally close-packed granular membrane. The black arrow indicates the direction of wave propagation. (b-d) Illustrations of longitudinal, shear, and rotational motions activating normal and shear nonlinear contact springs.}
		\label{Fig1}
	\end{figure}

	We consider a 2D, hexagonally close-packed lattice of spheres, as shown in Fig. \ref{Fig1}(a). The interactions between spheres follow the Derjaguin-Muller-Toporov (DMT) adhesion model \cite{DMT1983, Israelachvili}, which includes Hertzian normal contact forces \cite{Hertz,Johnson}, and a static adhesive force $ F_{DMT} = 2 \pi w R_c  $ due to van der Waals interactions, where $ w $ is the work of adhesion \cite{Israelachvili}, $R_c=R/2$ is the effective radius for two spheres in contact, and $R$ is the microsphere radius. To describe the shear contact interactions, we use the Hertz-Mindlin model \cite{Johnson,Mindlin}, assuming no slip occurs in the contact surface. Because the characteristic sound speeds of waves in the lattice are much slower than those of the bulk material of the spheres, we treat the spheres as rigid bodies with radius $ R $, mass $ m $, and moment of inertia $ I = (2/5) m R^2 $, interacting via nonlinear spring elements \cite{NesterenkoBook}. In terms of the displacements $ u_{j,k} $, $ v_{j,k} $, and $ \theta_{j,k} $, which represent horizontal, vertical, and angular displacements from equilibrium, respectively, the equations of motion of the sphere with index $ (j,k) $ are given by
	
	\begin{eqnarray}
	m \ddot{u}_{j,k} &=& f_N(\bar{\delta}_2) - f_N(\bar{\delta}_5) + \frac{1}{2} (f_N(\bar{\delta}_1) - f_N(\bar{\delta}_4) + f_N(\bar{\delta}_3) - f_N(\bar{\delta}_6))  \nonumber \\ &+& \sqrt{\frac{3}{2}}(f_S(\bar{\delta}_3) - f_S(\bar{\delta}_6) - f_S(\bar{\delta}_1) + f_S(\bar{\delta}_4)) \label{eq: equ} \\
	m \ddot{v}_{j,k} &=& f_S(\bar{\delta}_2) - f_S(\bar{\delta}_5) + \frac{1}{2} (f_S(\bar{\delta}_1) - f_S(\bar{\delta}_4) + f_S(\bar{\delta}_3) - f_S(\bar{\delta}_6))  \nonumber \\
	&+& \sqrt{\frac{3}{2}}(f_N(\bar{\delta}_1) - f_N(\bar{\delta}_4) - f_N(\bar{\delta}_3) + f_N(\bar{\delta}_6)) \label{eq: eqv}\\
	I \ddot{\theta}_{j,k} &=& R \sum_l f_S(\bar{\delta}_l),  \label{eq: eqphi}
	\end{eqnarray}
	
	\noindent
	where $ \bar{\delta}_l  = (\delta_{l,N}, \delta_{l,S})$ is the vector of relative normal and tangential displacements between particles labeled $ l $ and $ 0 $, as is defined in the Appendix. The normal spring force (positive away from particle $ 0 $)  and shear spring force (positive when inducing a counter-clockwise moment about particle $ 0 $) are given by 
	
	\begin{eqnarray}
	f_N(\bar{\delta}_l) &=& -\frac{4}{3} E^*R_c^{1/2}\left[\Delta_0 - \delta_{l,N}\right]_+^{3/2} \label{springN}\\
	f_S(\bar{\delta}_l) &=& 8 G^* R_c^{1/2} \delta_{l,S} \left[\Delta_0 - \delta_{l,N}\right]_+^{1/2}, \label{springS}
	\end{eqnarray}
	
	\noindent
	respectively. Here, $ \Delta_0 = [3 F_{DMT}/(4 E^* R_c^{1/2})]^{2/3} $ is the static overlap due to adhesion, and $ E^* = E/(2(1-\nu^2)) $  and $ G^* = G/(2(2-\nu)) $ are the effective elastic and shear moduli, respectively, of a solid with elastic modulus $ E $, shear modulus $ G $, and Poisson's ratio $ \nu $. As shown in Fig. \ref{Fig1}(b-d), the normal springs are activated by axial sphere displacements, while the shear springs can be activated by both transverse displacements and rotations.
	
	We expand the nonlinear spring forces (\ref{springN}) and (\ref{springS}) in Taylor series up to quadratic order as
	
	\begin{eqnarray}
	f_N(\bar{\delta}_l)  &\simeq& -A\Delta_0^{3/2} +  \frac{3}{2} A \Delta_0^{1/2} \delta_{l,N} - \frac{3}{8} A \Delta_0^{-1/2} \delta_{l,N}^2 + \mathcal{O}(\delta_{l,N}^3)\\
	f_S(\bar{\delta}_l) &\simeq& B \Delta_0^{1/2} \delta_{l,S} - \frac{1}{2} B \Delta_0^{-1/2} \delta_{l,S} \delta_{l,N} + \mathcal{O}(\delta_{l,S} \delta_{l,N}^2),
	\end{eqnarray}
	
	\noindent
	where $ A = (4/3) E^* R_c^{1/2} $ and $ B = 8 G^* R_c^{1/2} $. We name the linear and quadratic stiffnesses $ k_1 = (3/2) A \Delta_0^{1/2} $, $ g_1 =  B \Delta_0^{1/2} $, $ k_2 =  -(3/8) A \Delta_0^{-1/2}$, and $ g_2 =  -(1/2) B \Delta_0^{-1/2}$, and define the nonlinearity parameter $ \epsilon = (3/2) D_0 (g_2 + 3 k_2)/(g1 + 3 k_1) $. 
	
	We restrict our analysis to plane waves traveling in the positive $ y $ direction; hence, the equations of motion take the form of an effective 1D lattice with three degrees of freedom, and we adopt the single index $ n $. We also transform the system to dimensionless variables $ p = v/D_0 $, $ q = -u/D_0 $, $ \phi = R \theta /D_0 $, and $ \tau = t/(\sqrt{2m/(g_1+3 k_1)}) $, where $ D_0 $ is a characteristic displacement amplitude. Finally, the equations of motion are reduced to a simpler form, which is valid for small nonlinearity ($ \epsilon << 1 $):

	\begin{eqnarray}
	p^{''}_n &= & \frac{1}{4} \nabla_n^2 p + \epsilon [\frac{1}{8}(\nabla_n^2 p) (\nabla_n p) - \frac{1}{6} \alpha (\nabla_n^2 q) (\nabla_n q)  \nonumber\\
	&+& \frac{1}{3\sqrt{3}} \beta (\phi_n \nabla_n^2 q - q_n \nabla_n^+ \phi + \nabla_n^+(q\phi))] 
	\label{eqp}\\
	q^{''}_n &= & \frac{\mu_1}{4} \nabla_n^2 q - \frac{\sqrt{3}\mu_2}{2} \nabla_n \phi - \frac{\epsilon}{4}[\gamma (p_n \nabla_n q + q_n \nabla_n p - \nabla_n(qp))  \nonumber\\
	&+& \sqrt{3} \beta (-p_n \nabla_n^{2,+} \phi + \phi_n \nabla_n^+ p + \nabla_n^+ (\phi p))]
	\label{eqq}\\
	\phi^{''}_n &= & \frac{\mu_2}{2 \tilde{I}} [-2(\nabla_n^{2,+} \phi + 2 \phi_n) + \sqrt{3} \nabla_n q]   \nonumber\\
	&+& \epsilon \frac{\beta}{2 \tilde{I}}  [\frac{1}{\sqrt{3}} (-q_n \nabla_n^2 p - p_n \nabla_n^+ q + \nabla_n^+(qp)) + (p_n \nabla_n \phi - \phi_n \nabla_n p - \nabla_n(\phi p))], 
	\label{eqr}
	\end{eqnarray}

	\noindent
	where the $ (\cdotp)'' $ notation denotes the second derivative with respect to $ \tau $, the parameters $ \mu_1 = (3 g_1 + k_1)/(g_1 + 3 k_1) $, $  \mu_2 = g_1/(g_1 + 3 k_1)$, $ \alpha =  (g_2 - k_2)/(g_2 + 3 k_2)$, $ \beta = g_2/(g_2 + 3 k_2)  $, $ \gamma =  (g_2 + k_2)/(g_2 + 3 k_2) $, and $ \tilde{I} =  I / (m R^2) $, and the difference operators $ \nabla_n (\cdotp) = (\cdotp)_{i+1} - (\cdotp)_{i-1}  $, $ \nabla_n^2 (\cdotp) = (\cdotp)_{i+1} - 2 (\cdotp)_n + (\cdotp)_{i-1} $, $ \nabla_n^+ (\cdotp) =  (\cdotp)_{i+1} + (\cdotp)_{i-1}$, and $ \nabla_n^{2,+} (\cdotp) =  (\cdotp)_{i+1} + 2 (\cdotp)_n + (\cdotp)_{i-1}$. We note that the rescaled variables $ p_n $ and $ q_n $ represent displacements parallel and transverse to the direction of propagation, respectively.
	
	In all numerical results that follow, we use, as a case study, the geometric and material properties of $1$ $ \mu \mathrm{m} $ silica spheres: elastic constants \(E =\) $73$ GPa and \(\nu =\) $0.17$ \cite{GlassProp}, and work of adhesion \(w =\) $0.063$ J/m\textsuperscript{2} \cite{Israelachvili}.

	\subsection{Quasi-linear Regime}
	
\begin{figure}[h]
		\centering
		\includegraphics[width=0.75\linewidth]{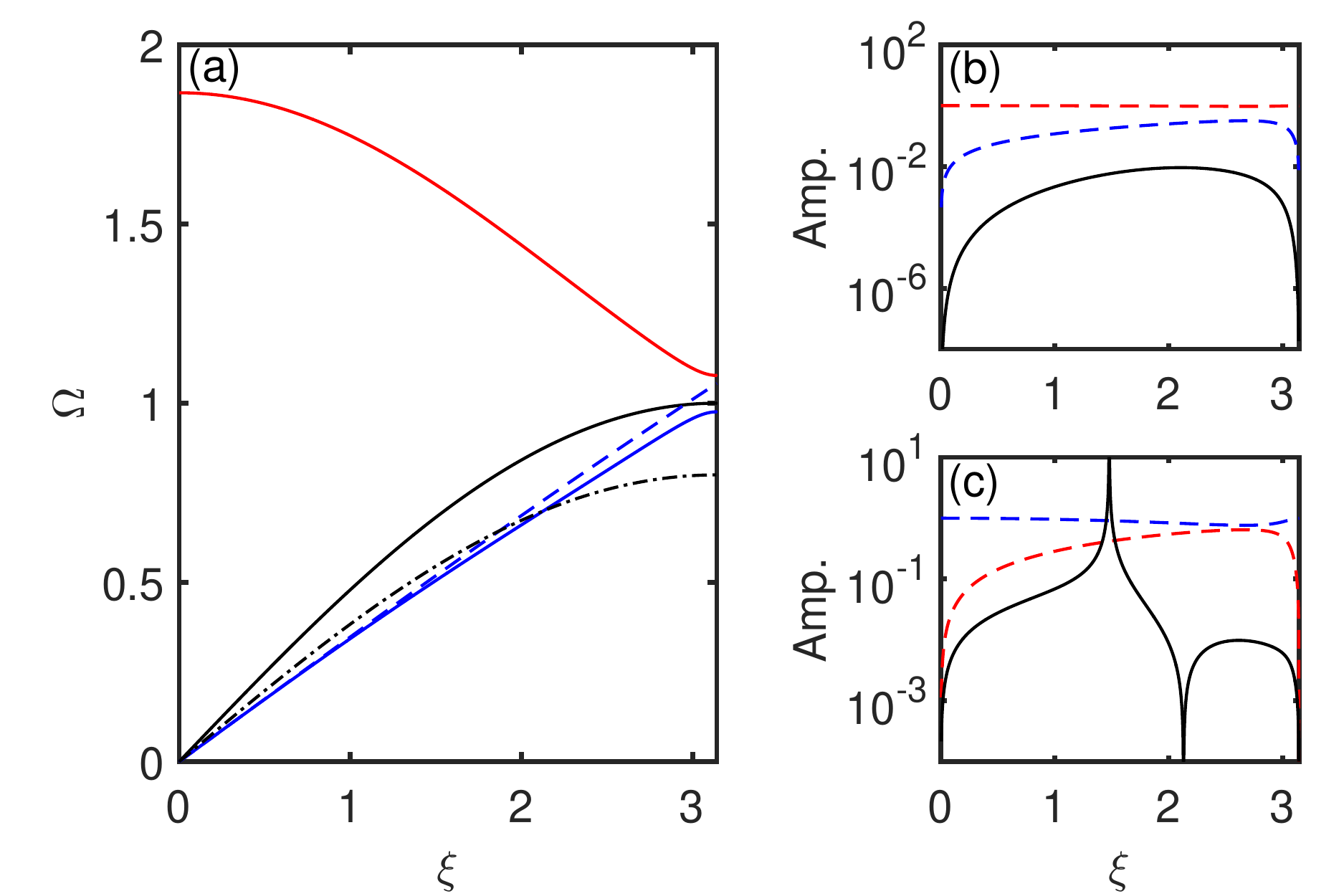}
		\caption{(a) Dispersion of plane waves traveling in the direction indicated in Fig. 1(a). Longitudinal (black), TR (blue), and RT (red) modes are denoted by the solid lines. The blue dashed curve is a frequency- and wave number-doubled representation of the TR mode, which is shown to intersect with the L mode. The black dash-dotted curve defines the antiresonance condition, wherein its intersection with the TR branch denotes the antiresonance frequency and wavenumber.  (b,c) Amplitudes of second harmonic longitudinal waves (black solid curves) generated from fundamental waves in the RT and TR modes, respectively. Blue and red dashed curves show the transverse and rotational displacements, normalized such that the sum of squares is unity. The longitudinal amplitudes are normalized by the small parameter $ \epsilon $.}
		\label{Fig2}
	\end{figure}

	In the limit of vanishing amplitude, the dynamics are linear, and the medium behaves as the 2D hexagonal lattice studied in \cite{Suiker2001}. Using Eqs. (\ref{eqp}) - (\ref{eqr}) with $ \epsilon = 0 $, we derive the dispersion relation, and find three modes, as shown in Fig. \ref{Fig2}(a): one that involves only longitudinal motion, and two involving coupled transverse and rotational motions \cite{Suiker2001, ZGVMembrane}. The dispersion equations are:
	
	\begin{eqnarray}
	\Omega &=& \sin{\left(\frac{\xi}{2}\right)} \label{Ldisp}\\
	\det{(M)} &=& 0 \label{TRdisp},
	\end{eqnarray}
	\noindent
	where $ \xi $ and $ \Omega $ are the dimensionless wave number and angular frequency (normalized to the lattice spacing in the $ y $ direction and longitudinal mode cutoff frequency \cite{BrillouinBook}, respectively), and the transverse-rotational dispersion matrix $ M $ is given by
	
	\begin{equation}
	M  = \left[ \begin{array}{cc}
	\frac{\mu_1}{2} [1-\cos(\xi)] - \Omega^2 & \sqrt{3} i \mu_2 \sin(\xi)\\
	-\sqrt{3} i \mu_2 \sin(\xi) & 2\mu_2 [2 + \cos(\xi)] - \tilde{I} \Omega^2
	\end{array}\right].
	\end{equation}
	
	Following conventions used in previous works \cite{Linear3DTheory, Linear3DExp}, we denote the longitudinal mode ``L,'' and the lower and upper transverse-rotational modes as ``TR'' and ``RT,'' respectively, where the first letter corresponds to the predominant displacement of each mode. While models for 3D hexagonally close-packed granular crystals \cite{Linear3DTheory}, which include shear interactions and particle rotations, predict faster wave speeds, due to additional interparticle coupling terms, and additional plane wave modes due to periodically alternating layers (including purely rotational modes), we suggest that our model may also be utilized to gain qualitative understanding of certain 3D scenarios. This is due to the similarity of acoustic-type L and TR modes, as well as one of the optical-type RT modes, present in both 2D and 3D systems.   
	
	\subsection{Weakly-nonlinear Regime: Second Harmonic Generation and Nonlinear Mode Conversion}
	For $ | \epsilon | << 1 $, the nonlinear response can be predicted using a successive-approximations approach, in the manner of Refs. \cite{Cabaret2012, Sanchez-Morcillo2013}. Using this approach, we represent the displacements as power series in $ \epsilon $:
	
	\begin{eqnarray}
	p_n(\tau) &=& P_{0,n}(\tau) + \epsilon P_{1,n}(\tau) +  \epsilon^2 P_{2,n}(\tau) + \dots \label{pseries}\\
	q_n(\tau) &=& Q_{0,n}(\tau) + \epsilon Q_{1,n}(\tau) +  \epsilon^2 Q_{2,n}(\tau) + \dots \label{qseries}\\
	\phi_n(\tau) &= &\Phi_{0,n}(\tau) + \epsilon \Phi_{1,n}(\tau) +  \epsilon^2 \Phi_{2,n}(\tau) + \dots \label{rseries},
	\end{eqnarray}
	\noindent
	and explore second harmonic generation in the granular medium, separately considering the cases of TR/RT and L modes as FF waves. While previous works employing this approach have focused on finite-length and semi-infinite lattices \cite{Cabaret2012, Sanchez-Morcillo2013} with a harmonic boundary condition, we consider an infinite lattice, and use a single plane wave of infinite extent that travels in the positive-$ n $ direction as the initial condition; that is, 
	
	\begin{eqnarray}
	p_n(0) &=& P_{0,n}(0) = \tilde{P}_0e^{i(\xi n)}, \label{IC}\\
	p'_n(0) &=& P'_{0,n}(0) = -i\Omega \tilde{P}_0e^{i(\xi n)}, \label{ICprime}
	\end{eqnarray}
	
	\noindent
	where $\tilde{P}_0$ is a constant amplitude, with similar expressions for the other displacements. Thus, the second harmonic response is a function of wave number. This configuration closely resembles the conditions of several recent microgranular experiments \cite{MicrogranularLaser} that utilize the laser-induced transient grating spectroscopy technique \cite{TG}, which excite waves with specific, defined wavelengths via the projection of a spatially periodic optical interference pattern. 
	
	We note that, for the expansions (\ref{pseries}) - (\ref{rseries}) to be valid, each function $ P_0$, $P_1 $, etc. should be of order $ \mathcal{O}(1) $; this imposes limits on the range of wave numbers for which the analysis is valid, as will be discussed later.

	\subsubsection{TR and RT Modes as Fundamental Waves}

	We first consider TR and RT modes as FF waves, i.e.
	
	\begin{eqnarray}
	p_n(0) &=& p'_n(0) = 0  \label{ICpTR} \\
	q_n(0) &=& Q_{0,n}(0) = \tilde{Q}_0e^{i(\xi n - \Omega \tau)}\rvert_{\tau = 0},  q'_n(0) = -i \Omega q_n(0) \label{ICqTR} \\
	\phi_n(0) &=& \Phi_{0,n}(0) = \tilde{\Phi}_0e^{i(\xi n - \Omega \tau)}\rvert_{\tau = 0}, \phi'_n(0) = -i \Omega \phi_n(0). \label{ICrTR}
	\end{eqnarray}
	\\
	\noindent
	We proceed by substituting the power series (\ref{pseries}) - (\ref{rseries}) into Eqs. (\ref{eqp}) - (\ref{eqr}).  At order $ \mathcal{O}(\epsilon^0) $, we find that $ P_{0,n} $, $ Q_{0,n} $, and $ \Phi_{0,n} $ must be plane waves satisfying the linear dispersion equations (\ref{Ldisp}) and (\ref{TRdisp}). To satisfy the initial condition (\ref{ICpTR}), it follows that $ P_{0,n}(\tau) = 0 $. Furthermore, the ratio of amplitudes of the $ Q_{0,n} $ and $ \Phi_{0,n} $ plane waves is constrained; in particular, a vector composed of these two amplitudes must be an eigenvector of the dispersion matrix $ M $ corresponding to the eigenvalues $ \xi $ and $ \Omega $. Because we have chosen a plane wave in the TR or RT mode as the initial condition, this constraint is trivially satisfied.
	
	At order $ \mathcal{O}(\epsilon^1) $, we find
	
	\begin{eqnarray}
	P''_{1,n} -  \frac{1}{4} (P_{1,n+1} - 2 P_{1,n} + P_{1,n-1}) &=& \frac{1}{8}(\nabla_n^2 P_0) (\nabla_n P_0) - \frac{1}{6} \alpha (\nabla_n^2 Q_0) (\nabla_n Q_0)   \nonumber\\
	&+& \frac{1}{3\sqrt{3}} \beta (\Phi_{0,n} \nabla_n^2 Q_0 - Q_{0,n} \nabla_n^+ \Phi_0 + \nabla_n^+(Q_0 \Phi_0))  \label{eqP1TR}\\
	Q''_{1,n}  -  \frac{\mu_1}{4} \nabla_n^2 Q_1 +  \frac{\sqrt{3} \mu_2}{2} \nabla_n \Phi_1 &=& - \frac{1}{4}[\gamma (P_{0,n} \nabla_n Q_0 + Q_{0,n} \nabla_n P_0 - \nabla_n (Q_0 P_0))  \nonumber\\
	&+& \frac{\sqrt{3}}{2} \beta (-P_{0,n} \nabla_n^{2,+} \Phi_0 + \Phi_{0,n} \nabla_n^+ P_0 + \nabla_n^+ (\Phi_0 P_0))] \label{eqQ1TR}\\
	\Phi''_{1,n} - \frac{\mu_2}{2 \tilde{I}} [-2(\nabla_n^{2,+}\Phi_1 + 2 \Phi_{1,n}) + \sqrt{3} \nabla_n Q_1]&= & \frac{\beta}{2 \tilde{I}}  [\frac{1}{\sqrt{3}} (-Q_{0,n}\nabla_n^2 P_0 - P_{0,n} \nabla_n^+ Q_0 + \nabla_n^+(Q_0 P_0))  \nonumber\\
	&+& (P_{0,n} \nabla_n \Phi_0 - \Phi_{0,n} \nabla_n P_0 - \nabla_n(\Phi_0 P_0))]. \label{eqR1TR}
	\end{eqnarray}
	\\
	\noindent
	Since $ P_{0,n} = 0 $, the right-hand side of Eq. (\ref{eqP1TR}) is composed of quadratic products of $ Q_{0,n} $ and $ \Phi_{0,n} $, while the right-hand sides of Eqs. (\ref{eqQ1TR}) and (\ref{eqR1TR}) vanish. Thus, the second harmonics at order $ \mathcal{O}(\epsilon^1) $ generated by fundamental plane waves in the TR and RT modes are purely longitudinal in character.  Furthermore, from the initial conditions (\ref{ICqTR}) and (\ref{ICrTR}), it follows that $ Q_{1,n}(\tau)  = \Phi_{1,n}(\tau) = 0$.
	
	We derive an expression for $ P_{1,n}(\tau)  $ by seeking a solution of the form
	
	\begin{equation}
	P_{1,n}(\tau) = \tilde{P}_{1,n}^{(+)} e^{i (2\xi n - \Omega(2\xi) \tau)} + \tilde{P}_{1,n}^{(-)} e^{i (2\xi n + \Omega(2\xi) \tau)} + \tilde{P}_{1,n}^{(TR)} e^{i (2\xi n - 2 \Omega \tau)}, \label{ansatzPTR}
	\end{equation}
	
	\noindent
	where $ \Omega(2 \xi) $ is found using Eq. (\ref{Ldisp}). The first two terms of ansatz (\ref{ansatzPTR}) are the homogeneous part of the solution, with the $ (+) $ and $ (-) $ superscripts corresponding to positive- and negative-$ n $ traveling waves, respectively. The third term is generated by the FF wave terms on the right-hand side of Eq. (\ref{eqP1TR}), and the superscript $ (TR) $ signifies that the generating FF wave is in either transverse-rotational mode. To derive an expression for the amplitude of the generated SH wave, we substitute into Eq. (\ref{eqP1TR}) the expressions $ P_{1,n} = \tilde{P}_{1,n}^{(TR)} e^{i (2\xi n - 2 \Omega \tau)}$, $ Q_{0,n} = \tilde{Q}_0e^{i(\xi n - \Omega \tau)}$ and $ \Phi_{0,n} = \tilde{\Phi}_0e^{i(\xi n - \Omega \tau)} $, and adopt the convention that the FF amplitudes $ \tilde{Q}_0 $ and $ \tilde{\Phi}_0 $ have been normalized, such that the sum of their squares is unity. After simplification, we find an expression for the amplitude of the longitudinal SH wave that depends on the amplitudes of the transverse-rotational FF waves, as well as the wave number:
	
	\begin{equation}
	\tilde{P}_{1,n}^{(TR)}(\xi) = \frac{\frac{4}{3} \sin(\xi)  \tilde{Q}_0 \left[ -\frac{\beta}{\sqrt{3}} \sin(\xi) \tilde{\Phi}_0 + i \alpha \sin^2(\xi/2) \tilde{Q}_0 \right]}{\sin^2(\xi) - (2\Omega)^2}, \label{P1TR}
	\end{equation}
	
	\noindent
	where the ratio $ \tilde{Q}_0 / \tilde{\Phi}_0 $ is a function of $ \xi $. This expression is applicable for FF waves in both the TR and RT modes, and is plotted for each case in Fig. \ref{Fig2}(b-c).
	
	In Fig. \ref{Fig2}(c), we observe a resonance near $ \xi \simeq \pi/2 $; this arises because the generated SH wave of the TR mode (i.e. the third term of (\ref{ansatzPTR})) has a wave number and frequency that intersect the L branch, as shown in Fig. \ref{Fig2}(a). Since the value of $ \tilde{P}_{1,n}^{(TR)} $ near the resonance is larger than $ \mathcal{O}(1) $, the expansion (\ref{pseries}) is no longer valid, as the generated harmonics would grow to the same order of magnitude as the FF wave. The wave number at which this resonance occurs depends on the mass and stiffness properties of the lattice, but the resonance must exist in our model, because the generated SH wave will always possess an intersection with the L mode, though it may occur outside the first Brillouin zone. In the limiting case of infinitely large rotational inertia or vanishing shear contact stiffness (i.e. neglecting rotations or shear interactions), the resonance still exists, albeit at a different wave number.

	We also note the presence of an antiresonance in Fig. \ref{Fig2}(c); this implies that the quadratic nonlinearities vanish at order $ \mathcal{O}(\epsilon^1) $, for a particular wavelength. Near the antiresonance, the harmonics generated by higher-order nonlinearities can be of the same order of magnitude as the SH waves; thus, these nonlinearities would need to be included in the equations of motion (\ref{eqp}) - (\ref{eqr}), and additional terms of the expansions (\ref{pseries}) - (\ref{rseries}) should be included, to obtain accurate results. The wave number of this antiresonance is dependent on mass and stiffness properties, but is not guaranteed to exist for arbitrary parameter values; indeed, the numerator of Eq. (\ref{P1TR}) is not guaranteed to vanish, except for values of $ \xi $ that are integer multiples of $ \pi $ (corresponding to the edges of the Brillouin zones). We note that the antiresonance cannot exist in the limit of infinitely large rotational inertia, i.e. $ \tilde{\Phi}_0 = 0 $; this underscores the importance of particle rotations in our model. 
	
	Considering finite rotational inertia, we derive a condition on the wave number $ \xi $ and frequency $ \omega $ that must be satisfied at the antiresonance. The dispersion relation (\ref{TRdisp}) is a solvability condition of the matrix equation
	
	\begin{equation}
	M \left[ \begin{array}{c}
	\tilde{Q}_0 \\
	\tilde{\Phi}_0	
	\end{array} \right]
	= \left[ \begin{array}{c}
	0\\
	0\end{array} \right].
	\end{equation} 
	
	\noindent
	Thus, by manipulating either of the two scalar equations in this system, the ratio $ \tilde{\Phi}_0 / \tilde{Q}_0$ can be written in terms of $ \xi $, $ \Omega $, and lattice stiffness and mass parameters. We find
	
	\begin{equation}
	\frac{\tilde{\Phi}_0}{\tilde{Q}_0} = - \frac{\mu_1 (1-\cos{(\xi)}) - \Omega^2}{2 \sqrt{3} i \mu_2 \sin{(\xi)}}. \label{PhiQratio}
	\end{equation}
	
	\noindent
	By substituting Eq. (\ref{PhiQratio}) into the bracketed term in the numerator of (\ref{P1TR}), equating this expression to zero, and rearranging, we derive the expression
	
	\begin{equation}
	\Omega = \sin{\left(\frac{\xi}{2}\right)}\sqrt{\mu_1 - 3 \mu_2\frac{\alpha}{\beta}}, \label{AntiRes}
	\end{equation}
	
	\noindent
	which defines a condition for the antiresonance. Thus, if the TR dispersion branch intersects the curve defined by Eq. (\ref{AntiRes}), the antiresonance exists at the corresponding wave number and frequency, as shown in Fig. \ref{Fig2}(a). Furthermore, when the Hertz-Mindlin contact model is assumed, the stiffness parameters contained in (\ref{AntiRes}) are all related by Poisson's ratio $ \nu $ \cite{Johnson}. By manipulating (\ref{TRdisp}) and (\ref{AntiRes}), it is straightforward to show that the TR mode has a higher cutoff frequency at $ \xi = \pi $ than the curve defined by (\ref{AntiRes}), but has a lesser slope at the origin. It follows that the intersection is guaranteed to exist for physically realizable materials (i.e. $ -1 \leq \nu \leq 0.5) $.
	
	Enforcing the initial conditions (\ref{IC}) and (\ref{ICprime}), which imply that $ P_{1,n}(0) = P'_{1,n}(0) = 0 $, we derive the amplitudes of the homogeneous terms as a function of the generated second harmonic amplitude:
	
	\begin{eqnarray}
	\tilde{P}_{1,n}^{(+)} &=& -\frac{\Omega(2\xi) + 2 \Omega}{2 \Omega(2\xi)}  \tilde{P}_{1,n}^{(TR)} \label{Pplus}\\
	\tilde{P}_{1,n}^{(-)} &=& -\frac{\Omega(2\xi) - 2 \Omega}{2 \Omega(2\xi)}  \tilde{P}_{1,n}^{(TR)}. \label{Pminus}
	\end{eqnarray}
	
	We note that, for an experimental setup, the FF wave would contain components traveling in both the positive and negative directions. In a region where these two components interact, two additional longitudinal SH waves would be present: a negative-$ n $ traveling wave analogous to the third term of (\ref{ansatzPTR}), which also has the same dependence on $ \xi $, and a standing wave with wave number $ 2\xi $ and vanishing frequency, which arises due to the interaction between FF waves.

	\subsubsection{L Mode as Fundamental Wave}

	We now conduct a similar analysis using a plane wave in the L mode as the FF wave, i.e. 
	
	\begin{eqnarray}
	p_n(0) &=& P_{0,n}(0) = \tilde{P}_0e^{i(\xi n - \Omega \tau)}\rvert_{\tau = 0}, p'_n(0) = -i\Omega p_n(0) \\
	q_n(0) &=& \phi_n(0) = 0\\
	q'_n(0) &=& \phi'_n(0) = 0.
	\end{eqnarray}
	\noindent
	By inspecting the quadratic terms in Eqs. (\ref{eqq}) - (\ref{eqr}), we observe that no motion in the transverse and rotational displacements can be generated via nonlinearity, because these terms are all products containing $ q_n $ and $ \phi_n $. Thus, the remaining equation of motion (\ref{eqp}) reduces to a form that is mathematically equivalent to that in \cite{Sanchez-Morcillo2013}, while the boundary conditions differ.
	
	After substituting the power series of (\ref{pseries}) into (\ref{eqp}), we equate the coefficients of order $ \mathcal{O}(\epsilon^0) $, seek solutions of the form $ P_{0,n}(\tau) = \tilde{P}_0e^{i(\xi n - \Omega \tau)} $, and find that $ P_{0,n}(\tau) $ satisfies (\ref{Ldisp}), the dispersion relation for L modes. At order $ \mathcal{O}(\epsilon^1) $, we find
	
	\begin{equation}
	P''_{1,n} -  \frac{1}{4} (P_{1,n+1} - 2 P_{1,n} + P_{1,n-1}) = \frac{1}{8}(P_{0,n+1} - 2 P_{0,n} + P_{0,n-1}) (P_{0,n+1} - P_{0,n-1}). \label{eqP1}
	\end{equation}
	
	\noindent
	This is a linear equation for $ P_{1,n} $ with a generation term that is quadratic in $ P_{0,n} $. Thus, we seek a solution as the sum of homogeneous and generated second harmonic waves in the form
	
	\begin{equation}
	P_{1,n}(\tau) = \tilde{P}_{1,n}^{(+)} e^{i (2\xi n - \Omega(2\xi) \tau)} + \tilde{P}_{1,n}^{(-)} e^{i (2\xi n + \Omega(2\xi) \tau)} + \tilde{P}_{1,n}^{(L)} e^{i (2\xi n - 2 \Omega \tau)}, \label{ansatzP}
	\end{equation}
	
	\noindent
	where each term is analogous to the corresponding term in Eq. (\ref{ansatzPTR}). Here, the superscript $ (L) $ signifies that the generating FF wave belongs to the longitudinal mode. The amplitude of the generated term $ \tilde{P}_{1,n}^{(L)} $, which is analogous to Eq. (\ref{P1TR}), can be derived as in \cite{Sanchez-Morcillo2013}, where a detailed analysis has been performed. While the system considered in \cite{Sanchez-Morcillo2013} is 1D and clearly not identical to our own, the calculation of $ \tilde{P}_{1,n}^{(L)} $ is mathematically equivalent.
	
	Because this solution must satisfy the initial conditions (\ref{IC}) and (\ref{ICprime}), the amplitudes of the homogeneous terms may by derived in the same way as for transverse-rotational FF waves, and appear in the same form as (\ref{Pplus}) and (\ref{Pminus}), with the exception that the superscript $ (TR) $ must be replaced by $ (L) $.

	\section{Simulations}
	
	\subsection{Numerical Setup}

	To examine the validity of our theoretical analysis with a complete Hertzian nonlinearity, we simulate second harmonic generation from an FF wave in the TR mode by numerically solving the equations of motion (\ref{eq: equ}) - (\ref{eq: eqphi}) for $ 1 < n < 201 $. We use initial conditions of the form (\ref{ICqTR}) - (\ref{ICrTR}), with fixed boundaries.  For each case, we probe the response of the central lattice site ($ n = 101 $), and end the simulation before disturbances caused by the boundaries reach this site. We present the results in terms of the normalized variables $ p_{101}(\tau) $, $ q_{101}(\tau) $, and $ \phi_{101}(\tau) $.

	\subsection{Numerical Results}

	\begin{figure}[h]
		\centering
		\includegraphics[width=\linewidth]{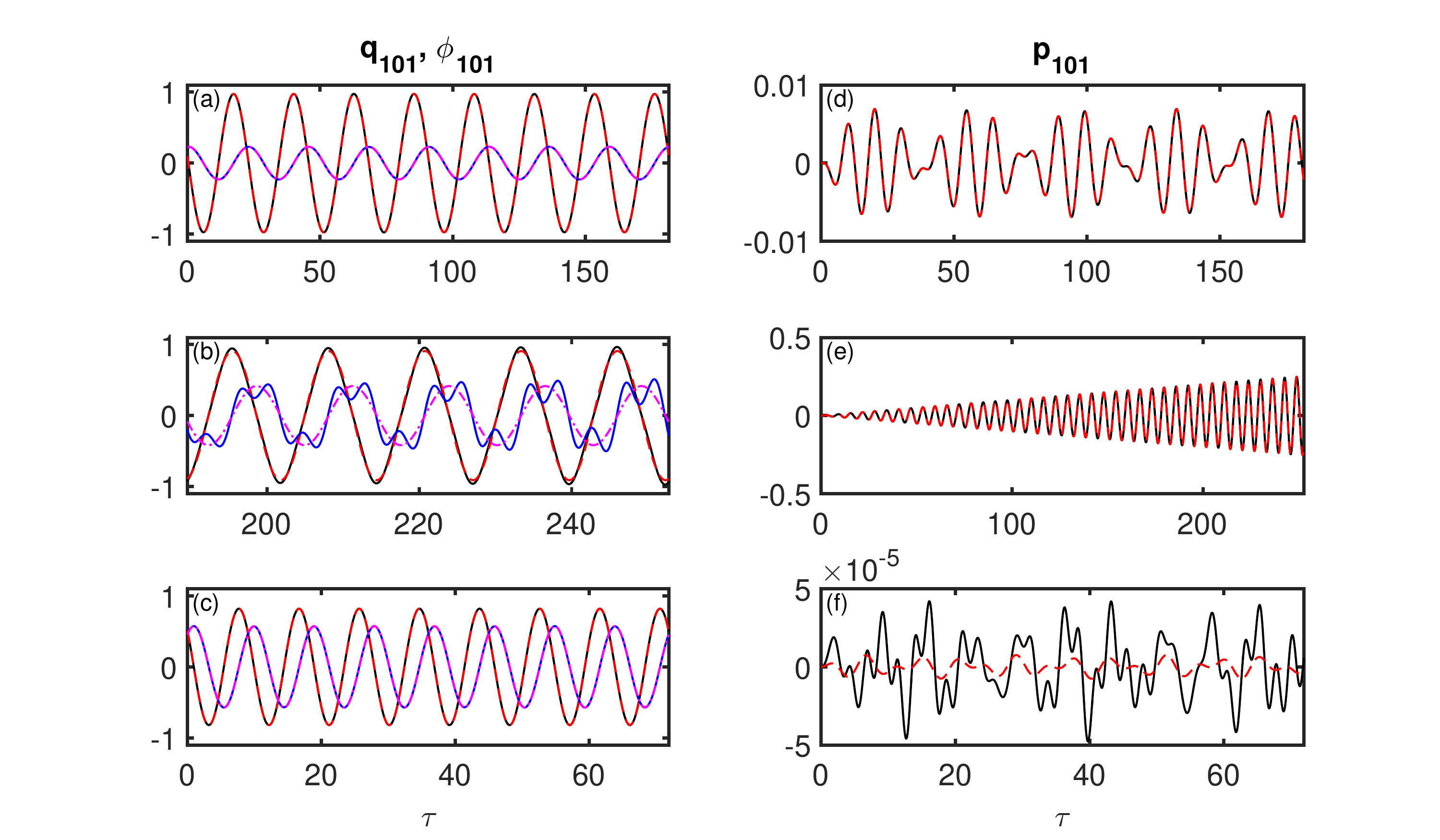}
		\caption{(a-c) Transverse and rotational displacements for $  \xi_1 = 0.80 $, $ \xi_2 = 1.475 $, and  $  \xi_3 = 2.127 $. Black and blue solid lines correspond to simulated transverse and rotational displacements, while red dashed and magenta dash-dotted lines show the corresponding theoretical predictions. (d-f) Simulated (black solid) and theoretical (red dashed) curves showing longitudinal displacements for the wave numbers of (a-c).}
		\label{Fig3}
	\end{figure}
	
	As shown in Fig. (\ref{Fig3}), we simulate the lattice using three wave numbers, each representing a qualitatively different case: 1) $ \xi_1 = 0.802 $, which is far from the resonance and antiresonance; 2) $ \xi_2 = 1.475 $, which is approximately the resonant wave number; and 3) $ \xi_3 =  2.127 $, which is approximately the antiresonant wave number. We have chosen the characteristic amplitude $ D_0 = 0.2 \Delta_0 $, which results in the nonlinearity parameter $ \epsilon_{sim} \approx -0.1 $.
	
	For case 1), the perturbation analysis is valid (note from Fig. \ref{Fig2}(c) that $ \tilde{P}_{1,n}^{(TR)} \simeq \mathcal{O}(1) $), and the simulated time histories are well-approximated by the theoretical predictions, as shown in Fig. \ref{Fig3}(a,d). In Fig. \ref{Fig3}(d), it can be seen that the SH amplitude exhibits ``beating'' at a frequency $ \Omega_{beat} = \Omega(2\xi) - 2\Omega $; this phenomenon is a well-known characteristic of SH generation \cite{NLOptics, Sanchez-Morcillo2013}. For case 2), the oscillations of $ p_{101} $ grow linearly in $ \tau $, as shown in Fig. \ref{Fig3}(e). While the theoretical prediction for $ p_{101} $ appears to match the simulation reasonably well, it is evident from Fig. \ref{Fig3}(b) that there is significant error in $ q_{101} $ and $ \phi_{101} $; in fact, the second harmonic is visibly present in the simulated signals. This is because, while the approximation works well for small $ \tau $, the generated SH longitudinal wave eventually reaches the same order of magnitude as the FF wave, thereby violating the assumption of weak nonlinearity. For case 3), the predictions for $ q_{101} $ and $ \phi_{101} $ are accurate, as shown in Fig. \ref{Fig3}(b). However, as shown in Fig. \ref{Fig3}(f), there is large error in $ p_{101} $, the generated longitudinal wave. This is because the harmonics generated by higher-order nonlinearities are larger than those coming from order $ \mathcal{O}(\epsilon^1 ) $; such harmonics are clearly visible in the simulated signal.
	
	\section{Conclusion}

	In this work, we have analyzed shear to longitudinal mode conversion via SH generation for plane waves in a 2D, adhesive, hexagonally close-packed microscale granular crystal, accounting for translational and rotational degrees of freedom, as well as normal and shear contact interactions. In the case of a FF wave in the L mode, we show that the generated SH is also longitudinal, and can be described using the same equations as for a 1D chain studied in earlier works. For the case where the lowest TR mode is treated as the FF wave, the generated SH wave is longitudinal, and an analytical expression for the SH amplitude reveals the presence of resonant and antiresonant wave numbers. This antiresonance is not predicted if rotations are excluded from the model, which demonstrates that, while the TR mode is predominantly transverse for most wavelengths, particle rotations produce a qualitative change in the nonlinear lattice dynamics. By simulating a lattice with DMT and Hertz-Mindlin particle interactions, we verify the accuracy of the theoretical analysis for a non-resonant case, and find that it effectively predicts the qualitative behavior of the resonant and antiresonant cases.  Future generalizations of this work to off-symmetry directions of propagation and extension to strongly nonlinear dynamical regimes may yield particularly interesting phenomena. This work is particularly suitable for the analysis of both microscale granular crystals, which are naturally precompressed by adhesive forces, and statically compressed granular crystals with macroscale particles. We expect that future studies of such 2- and 3D granular media will yield unique dynamics stemming from the interplay of interparticle shear coupling and particle rotations with highly nonlinear phenomena such as solitary waves \cite{NesterenkoBook} or hysteretic contact mechanics \cite{Tawfick}. 
	
	\section{Acknowledgments}
	This work was supported by the Army Research Office through grant no. W911NF-15-1-0030. 
	
	\appendix
	\section{Contact Spring Displacements}
	The normal component of the relative displacement between the particles labeled $ 0 $ and $ l $, denoted $ \delta_{l,N} $, is considered positive with increasing distance between the particle centers. The normal relative displacements are given by
	\begin{eqnarray}
	\delta_{1,N} &=& \frac{u_1 - u_0}{2} + \frac{\sqrt{3}(v_1-v_0)}{2}\nonumber \\
	\delta_{2,N} &=& u_2 - u_0\nonumber \\
	\delta_{3,N} &=& \frac{u_3 - u_0}{2} - \frac{\sqrt{3}(v_3-v_0)}{2}\nonumber \\
	\delta_{4,N} &=& -\frac{u_4 - u_0}{2} - \frac{\sqrt{3}(v_4-v_0)}{2}	\nonumber \\
	\delta_{5,N} &=& -(u_5 - u_0)\nonumber \\
	\delta_{6,N} &=& -\frac{u_6 - u_0}{2} + \frac{\sqrt{3}(v_6-v_0)}{2}.\nonumber 
	\end{eqnarray}
	
	The tangential component of the relative displacement, denoted $ \delta_{l,S} $, is considered positive if the shear force $ f_S(\bar{\delta}) $ (which is proportional to $ \delta_{l,S} $) induces a counterclockwise moment about particle $ 0 $. The tangential displacements are given by
	\begin{eqnarray}
	\delta_{1,S} &=& \frac{-\sqrt{3}(u_1 - u_0)}{2} + \frac{v_1-v_0}{2} - R(\theta_1 + \theta_0)\nonumber \\
	\delta_{2,S} &=& v_2 - v_0 - R(\theta_2 + \theta_0)\nonumber \\
	\delta_{3,S} &=& \frac{\sqrt{3}(u_3 - u_0)}{2} + \frac{v_3-v_0}{2} - R(\theta_3 + \theta_0)\nonumber \\
	\delta_{4,S} &=& \frac{\sqrt{3}(u_4 - u_0)}{2} - \frac{v_4-v_0}{2} - R(\theta_4 + \theta_0)	\nonumber \\
	\delta_{5,S} &=& -(v_5 - v_0) - R(\theta_5 + \theta_0)\nonumber \\
	\delta_{6,S} &=& -\frac{\sqrt{3}(u_6 - u_0)}{2} - \frac{v_6-v_6}{2} - R(\theta_6 + \theta_0).\nonumber 
	\end{eqnarray}

\end{document}